\documentstyle[psbox]{WORKSHOP}
\begin{document}

\def\pn{\par\noindent}
\def\chandra{{\it Chandra}}
\def\xmm{XMM--{\it Newton}}
\def\cgs{erg cm$^{-2}$ s$^{-1}$} 
\def\lum{erg s$^{-1}$} 
\def\cm{cm$^{-2}$} 
\def\gsimeq{\hbox{\raise0.5ex\hbox{$>\lower1.06ex\hbox{$\kern-1.07em{\sim}$}$}}} 
\def\lsimeq{\hbox{\raise0.5ex\hbox{$<\lower1.06ex\hbox{$\kern-1.07em{\sim}$}$}}} 

\title{
AGN unified scheme and evolution: a {\it Suzaku} view \\ 
}

\author{
Andrea Comastri$^1$, Kazushi Iwasawa$^1$, Roberto Gilli$^1$, \\
Cristian Vignali$^2$ and Piero Ranalli$^2$ 
\\[12pt]  
%
$^1$  INAF Osservatorio Astronomico di Bologna, via Ranzani 1, 40127 
Bologna, Italy\\
$^2$  Dipartimento di Astronomia, Universit\`a di Bologna, via Ranzani 1, 
40127 Bologna, Italy\\ 
%
{\it E-mail(AC): andrea.comastri@oabo.inaf.it} 
}

\abst{
   We present broad band {\it Suzaku} observations of a small sample of hard X--ray selected 
   ($>$ 10 keV), nearby Seyfert 2 galaxies and discuss the results in the context of AGN 
    unified model. We also review the issues related to the space density of 
    heavily obscured, Compton Thick AGN
   in the local Universe and the perspectives for the search of these objects at high redshifts.
}

\kword{X-rays: active galaxies --- X-rays: spectroscopy}

\maketitle
\thispagestyle{empty}

\section{Introduction}

Since the discovery of polarized broad emission lines in the 
archetypal Seyfert 2 galaxy NGC 1068 (Antonucci \& Miller 1985), 
the AGN Unified Model (UM) 
was tested over the entire electromagnetic spectrum.
While it has been realized that its original 
formulation was too simple to explain the 
large body of observational data, it still 
provides a useful framework for AGN studies.

Hard X--ray observations of Seyfert galaxies 
have shown that absorption by circumnuclear gas 
and dust, presumably with a toroidal geometry, is almost ubiquitous 
among optically classified Type 2 objects. 
Deep and wide X--ray surveys, combined with pointed observations of nearby 
bright AGN, allowed to probe a large interval 
of luminosities, redshifts and column densities and thus to test 
the AGN UM in the X--ray band.

A wide range of absorbing column densities is observed among Seyfert 2
and is required to fit the X--ray background ({\sc xrb}) spectrum (Gilli et al 2007; Treister 
et al. 2009). Absorption is much less common at high X--ray luminosities 
with a trend similar to that observed at optical and infrared wavelengths
(Simpson 2005, Maiolino et al. 2007).
There are also hints of an increasing fraction of obscured AGN 
towards high redshifts (La Franca et al. 2005; Treister et al. 2006). 
Besides providing additional tests to the UM, the above trends are most likely 
closely related to the growth and evolution of Supermassive Black Holes (SMBHs).
According to current models (e.g. Hopkins et al. 2008)
all the galaxies undergo a phase of heavy, possibly Compton Thick 
($N_H > 10^{24}$ cm$^{-2}$) obscuration, strong  accretion 
and star formation. The census of obscured AGN may thus provide 
useful insights on our understanding of a key phase of 
SMBH and galaxy co-evolution. 
Due to the lack of sensitive X--ray observations above 10 keV, Compton 
Thick AGN can be efficiently detected and recognized as such only in the 
local Universe.

In the following we present {\it Suzaku} observations of a small sample 
of hard X--ray selected Seyfert 2. The goal is to characterize the 
physics and geometry of the obscuring gas and, a first step, to 
investigate their local space density. We also briefly discuss the  
search for the most obscured sources in various X--ray surveys 
and the implications for the study of their properties at high redshifts.

\section{Suzaku observations of nearby Seyfert 2 galaxies}

We have conceived a program with {\it Suzaku} (Mitsuda et al. 2007) to 
observe five nearby, relatively X--ray bright ($> 10^{-11}$ \cgs) 
AGN. The sources were selected from the {\sc integral/ibis} 
(Beckmann et al. 2006) and {\sc swift/bat} (Markwardt et al. 2005) catalogues. 
The column densities, as inferred from archival {\it Chandra} 
and XMM--{\it Newton} observations, are of the order of $10^{23-24}$ cm$^{-2}$, 
though affected by large errors.
For all the sources in the sample, a significant detection is achieved 
with the hard X--ray detector up to 40--50 keV along with a good quality X--ray spectrum 
with the {\sc xis ccd} below 10 keV (Fig.~1 and Comastri et al. 2009). 
The high energy ($>$ 2-3 keV) spectra are fitted with an absorbed  power law plus 
a reflaction component and an iron line.

\begin{figure}[t]
\centering
\psbox[xsize=8cm]
{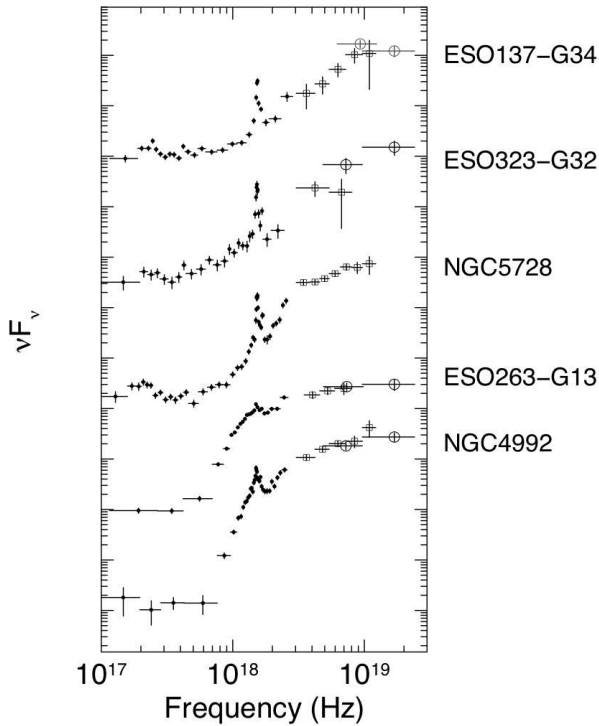}
\caption{The Suzaku spectra of the Seyfert 2 galaxy sample. The open circles at 
high energies are from Integral.} 
\end{figure}

Not surprisingly, heavy absorption with column densities in excess
of $10^{23}$ cm$^{-2}$ is measured in all the sources. Three of them are 
Compton Thick ($N_H > 10^{24}$ cm$^{-2}$) and among those, two are best 
fitted by a reflection dominated spectrum.
A summary of the spectral 
fit parameters, relevant for the present discussion, is reported in Table 1.
Soft X--ray emission, in excess of that expected by an extrapolation 
to lower energy of the absorbed spectrum, is clearly observed in all 
objects with only one exception.   
The low energy spectrum is fitted with two different models: (i) a 
partial covering (leaky absorber) model and (ii) a power law 
plus narrow Gaussian lines to approximate the emission of circumnuclear 
gas photoionized by the central nucleus. 

\subsection{Partial covering model}

The fits with a partial covering model, which provides a good description 
of the broad band spectra, allow us to compute the fraction ($f$) of 
primary X--ray emission scattered into the line of sight and related to the 
covering fraction  of the absorbing matter (e.g. the postulated torus 
in AGN UM). Interesting enough, the measured values 
of $f$ and the intensity of the reflection component ($R$) follow the correlation 
suggested by Eguchi et al. (2009) based on the analysis of a small sample 
of 6 Seyfert 2 galaxies detected by {\sc Swift/Bat} and selected in a very similar
way (Fig.~2). 
Following Eguchi et al (2009), sources with a low scattering fraction, dubbed ``New Type" 
AGN, are associated to a geometrically and optically thick configuration of the obscuring gas
seen rather face on. On the basis of a solid angle argument, they predict 
a large population of heavily Compton Thick 
AGN ($N_H > 10^{25}$ cm$^{-2}$) with extremely low scattering fraction 
which would remain largely undetected even in hard X--ray observations.
While it may be premature to invoke the presence of a new population on the 
basis of the present data, the existence of fully covered, heavily obscured AGN 
would have important consequences for the census of SMBHs.

\begin{figure}[t]
\centering
\psbox[xsize=7cm]
{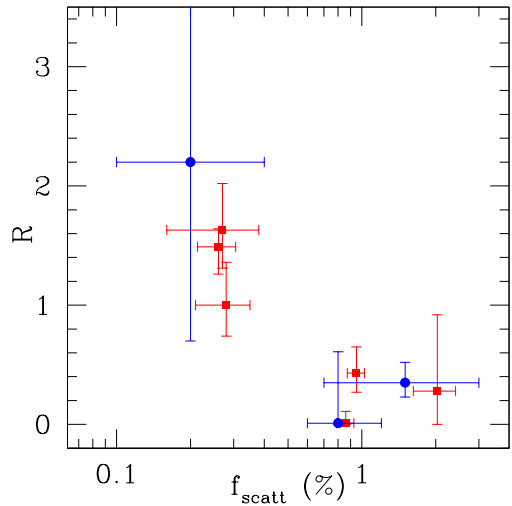}
\caption{Scattering fraction vs reflection component intensity. Red points are
from Eguchi et al. (2009), blue points refer to the 3 objects in the present sample 
for which the scattering fraction could be estimated.} 
\end{figure}

\begin{table}[h]
\caption{Suzaku sample }
\begin{center}
\begin{tabular}{lcccc} \hline\hline\\[-6pt]
Source Name & $N_H$ (cm$^{-2})$ & $f$(\%) & $R$   & K$\alpha$ EW (eV)\\   \hline
ESO137-G34  & 10$^{25}$ &       ...     &  ...    & 1350      \\
ESO323-G32  & 10$^{25}$ &       ...     &  ...    & 2200     \\
NGC 5728    & 1.5 10$^{24}$ &   1.5     &  0.35   & 1040     \\
ESO263-G13  & 2.6 10$^{23}$ &   0.8     &  $<$0.6 &   80     \\
NGC 4992    & 5.5 10$^{23}$ &   0.2     &  2.2    & 450      \\   \hline 
\end{tabular}
\end{center}
\end{table}

\subsection{Soft X--ray emission lines}

An equally good description in terms of spectral fits quality is obtained 
assuming that the soft X--ray emission is due to a blend of emission lines
which are tipically resolved in good signal to noise 
XMM--{\it Newton} reflection grating spectra (Guainazzi \& Bianchi 2007). 
While photoionization codes (e.g. {\sc xstar}) are usually adopted to model high resolution
soft X--ray spectra, the statistical quality  of the available  {\sc ccd} 
data is such that a simple parameterization, in terms of a power law plus narrow Gaussian 
lines, is well suited for the present purposes. 
The power law slope is free to vary and is independent from that of the obscured
high energy continuum to possibly account for ``residual" emission 
from thermal gas. Individual, narrow Gaussian 
lines are then added at the best fit energy 
of the most common transitions observed in other Seyfert galaxies.
The line energy and intensity are free parameters of the fit.
The number of lines considered for each source depends 
from the actual counting statistics. The ionized Oxygen 
{\sc oviii} line at 0.65 keV and {\sc neix} at 0.92 keV are present in
all the objects. A large fraction of the soft X--ray flux in NGC 4992, 
the source with the lowest scattering fraction in our sample (Fig.~3), can be accounted for 
in terms of line emission. The richest line spectrum is that of NGC 5728 
where nine indipendent lines (from Oxygen to Calcium) can be fitted (Fig.~4).

\begin{figure}[t]
\centering
\psbox[xsize=8cm,rotate=r]
{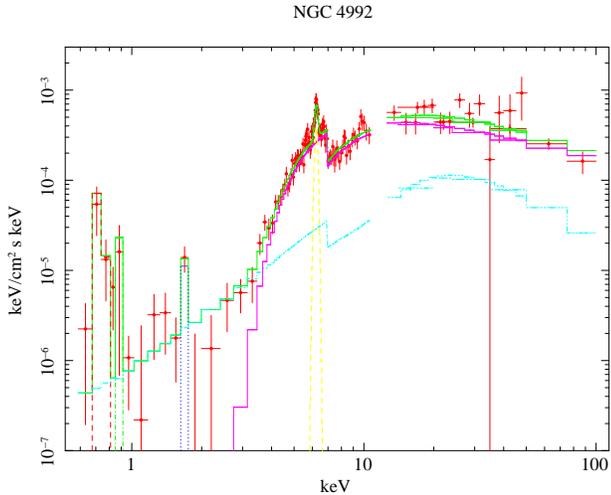}
\caption{The unfolded spectrum of NGC 4992, fitted with an absorbed power 
law plus a Iron line at high ($>$ 3--4 keV) energies. The weak soft X--ray spectrum 
can be fitted with a power law and a sum of Gaussian lines.}
\end{figure}

The best fit parameters of the hard X--ray continuum are fully consistent 
with those obtained with a partial covering fit.

\begin{figure}[t]
\centering
\psbox[xsize=8cm,rotate=r]
{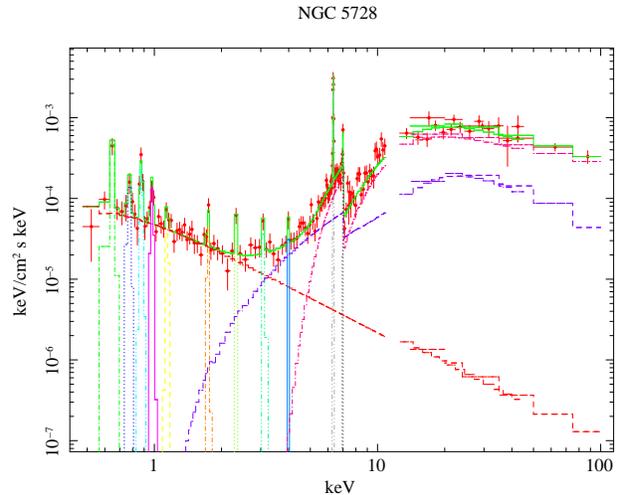}
\caption{The unfolded spectrum of NGC 5728, fitted with the same model adopted 
for NGC 4992. The nuclear source is obscured by  Compton Thick  gas and 
strong K$\alpha$ and K$\beta$ iron lines are detected} 
\end{figure}

Both models provide a good description of the observed spectra and 
it is not possible, with the available counting statistics, to choose a
``best fit" model. The explanation of the soft X--ray excess as 
a blend of unresolved (at the {\sc ccd} resolution) lines from a photoionized
plasma is consistent with previous X--ray observations 
of nearby, obscured AGN (Guainazzi \& Bianchi 2007).

\section{The space density of obscured AGN}

Hard X--ray selection is, in principle, almost unbiased against heavy obscuration
and thus considered to be well suited to estimate the intrinsic 
absorption distribution of AGN and, in particular, the relative 
fraction of Compton Thick  AGN in the local Universe.
The large majority of {\sc integral/ibis} and {\sc swift/bat} 
flux limited samples of bright AGN were observed by XMM and {\it Chandra} 
and were the subject of follow--up dedicated programs with {\it Suzaku}.
Surprisingly, only the already known Compton Thick AGN were recovered 
by the above mentioned hard X--ray surveys and no examples 
of newly discovered Compton Thick sources are reported in the literature. 
As a consequence, the relative fraction of Compton Thick AGN falls 
short by a factor of about 2 than that predicted by the Gilli et al. (2007)
XRB synthesis model at the {\sc integral} and {\sc swift} limiting 
fluxes. 
A relatively low contribution of CT AGN is predicted by the XRB 
synthesis model of Treister et al. (2009). By fixing the low redshift 
CT AGN fraction to that observed by hard X--ray surveys, they conclude that 
the total contribution of Compton Thick obscured accretion to the XRB is 
of the order of 10\%, to be compared with about 25\% of Gilli et al. (2007).
\pn 
It is important to stress that the observed fraction in the local Universe
is estimated using a relatively low number of sources and indeed the associated errors
are large. Nevertheless, it seems that present observations are favouring 
a relatively low space density of heavily obscured sources.
It is possible to reduce the discrepancy between the 
the original Gilli et al. (2007) model predictions and the observations
assuming a lower ratio (0.3 instead of 1) between obscured and unobscured, luminous 
($L_X > 10^{44}$ erg s$^{-1}$) AGN (upper envelope 
of the shaded region in figure 5). A better agreement (lower envelope of the shaded region in 
figure 5) is obtained assuming a slightly 
different $N_H$ distribution. More specifically, the space density of transmission 
dominated Compton thick AGN, detectable by surveys above 10 keV, is reduced and, at the 
same time, that of Compton thin and reflection dominated sources is increased
in such a way to keep approximately constant the total number of obscured AGN 
in the model and to fit the XRB spectrum. 

\begin{figure}[t]
\centering
\psbox[xsize=7cm]
{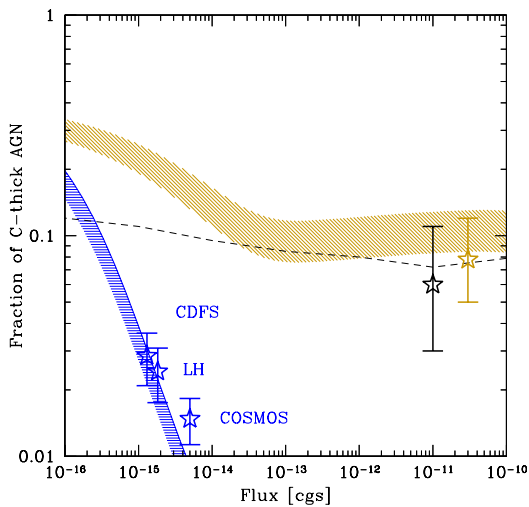}
\caption{ The fraction of Compton thick AGN as a function of the X--ray flux. The dashed line and the 
yellow shaded region correspond to the predictions of Treister et al. (2009) and Gilli et al. (2007)
respectively for the 20--40 keV band, while the points with error bars are the observational 
estimates from
BAT and IBIS. The blue shaded region and the points at low X--ray fluxes correspond to the Gilli et al 
predictions in the 2--10 keV band and the ``observed" fraction in various surveys as labeled, 
respectively.} 
\end{figure}

\pn 
The discrepancy between the two models is now negligible at bright 
X--ray fluxes, while it remains significant at lower fluxes and higher 
redshifts. 

Deep {\it Chandra} surveys have unveiled several examples of heavily obscured
candidate Compton Thick AGN (Tozzi et al. 2006; Georgantopoulos et al. 2009).
At the face value the observed fraction is consistent with the Gilli et al. (2007)
predictions in the 2--10 keV band, though the statistical fluctuations, due to the 
low number of objects and systematics errors associated to the column density determination,  
may well be larger than those plotted in figure 5.  

A significant improvement in the study of Compton thick AGN at $z\sim$ 1, and 
possibly beyond is expected by the ongoing ultra--deep (2.5 Msec) XMM--{\it Newton} 
survey in the CDFS.
Thanks to the large throughput of XMM it will be possible to obtain good quality 
X--ray spectra for a sizable number of sources.
The {\it pn} spectrum of a candidate CT AGN at $z=1.53$ in the
Tozzi et al. (2006) sample,  
obtained with about half of the total final exposure, is reported in figure 6. 
A strong (EW of about 1 keV) K$\alpha$ line is detected on top of a very 
hard continuum and is interpreted as the signature of Compton Thick obscuration.

Indirect searches for Compton Thick AGN at $z \simeq$ 2, using a mid--IR optical 
color selection (Daddi et al. 2007; Fiore 
et al. 2008; 2009), seem to suggest that heavily obscured accretion 
at high redshift may be common, in line with the theoretical expectations.
However the possible contamination from starbust galaxies is a serious issue
and the determination of the space density of Compton thick AGN at high--$z$ 
is still affected by large uncertainities. Forthcoming {\sc herschel} observations
will allow to disentangle starburst from nuclear emission, extending 
previous studies with {\sc spitzer} to higher redshifts.

The direct detection of high redshift, heavily obscured Compton Thick AGN 
and the statistical study of their properties (luminosity function and evolution) 
cannot be achieved without imaging observations in the hard X--rays.
In this respect, a major step forward is expected by future planned missions
such as {\sc astro-h} and {\sc nustar}.

\begin{figure}[t]
\centering
\psbox[xsize=8cm]
{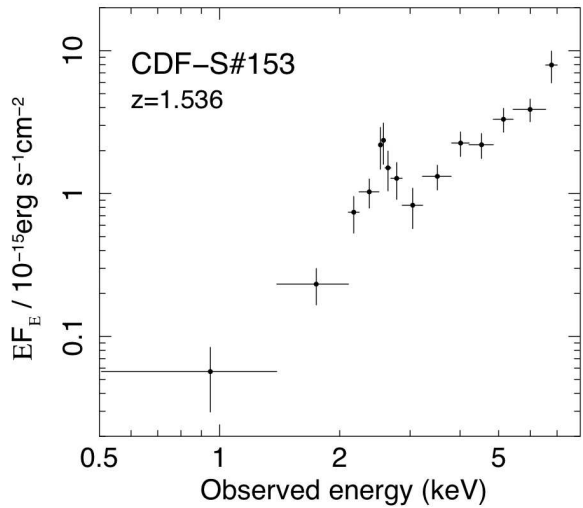}
\caption{The XMM-Newton spectrum of a candidate 
Compton Thick AGN in the CDFS (for a total exposure time of $\sim$ 1.5 Ms).}
\end{figure}

\vspace{1pc}

\section*{References}

\re
Antonucci R.R.J. \& Miller J.S. 1985 ApJ. 297, 621

\re
Beckmann V. et al. 2006 ApJ. 638, 642 

\re 
Comastri A. et al. 2009 ApJ. submitted 

\re
Daddi E. et al. 2007 ApJ. 670, 173 

\re
Eguchi S. et al. 2009 ApJ. 696, 1657 

\re
Fiore F. et al. 2008  ApJ. 672,, 94

\re
Fiore F. et al. 2009  ApJ. 693, 447 

\re
Georgantopoulos I., et al. 2009 arXiv0909.0224 

\re
Gilli R. et al. 2007 AA 463, 79 

\re
Guainazzi M. \& Bianchi S. 2007 MNRAS 374, 1290  

\re
Hopkins P.F. et al. 2008 ApJS. 175, 356 

\re
La Franca F. et al. 2005 ApJ. 635, 864 

\re
Maiolino R. et al. 2007 AA 468, 979 

\re
Markwardt et al. 2005 ApJ. 633, L77 

\re
Simpson C. 2005 MNRAS 360, 565 

\re
Tozzi P. et al. 2006 AA 451, 457 

\re
Treister E. \& Urry C.M. 2006 ApJ. 652, 79 

\re
Treister E. et al. 2009 ApJ. 696, 110 

\label{last}

\end{document}